\documentclass[aps,prl,twocolumn,superscriptaddress,calc,epsfig,10pt,longbibliography,floatfix]{revtex4-1}

\usepackage{graphicx}
\usepackage{graphics}
\usepackage{amssymb,amsmath}
\usepackage{wasysym}
\usepackage{float}
\usepackage{physics}
\usepackage{color}
\usepackage[normalem]{ulem} 
\usepackage[pdftex,breaklinks,colorlinks=true,linkcolor=blue, citecolor=blue,urlcolor=blue]{hyperref}

\newcommand{\mitCUAaddress}{Department of Physics, MIT-Harvard Center for Ultracold Atoms, and Research Laboratory of Electronics, MIT, Cambridge, Massachusetts 02139, USA}

\AtBeginDocument{\let\oldcontentsline\contentsline}
\newcommand{\notoccontentsline}[4]{\oldcontentsline{}{}{}{}}
\newcommand{\droptocpage}{\addtocontents{toc}{\let\protect\contentsline\protect\notoccontentsline}}
\newcommand{\incltocpage}{\addtocontents{toc}{\let\protect\contentsline\protect\oldcontentsline}}

\begin{document}

\title{
Doublon-hole correlations and fluctuation thermometry in a Fermi-Hubbard gas
}

\author{Thomas Hartke, Botond Oreg, Ningyuan Jia, and Martin Zwierlein}
\date{\today}

\affiliation{\mitCUAaddress}

\begin{abstract}
We report on the single atom and single site-resolved detection of the total density in a cold atom realization of the 2D Fermi-Hubbard model. Fluorescence imaging of doublons is achieved by splitting each lattice site into a double well, thereby separating atom pairs. Full density readout yields a direct measurement of the equation of state, including direct thermometry via the fluctuation-dissipation theorem. Site-resolved density correlations reveal the Pauli hole at low filling, and strong doublon-hole correlations near half filling. These are shown to account for the difference between local and non-local density fluctuations in the Mott insulator. Our technique enables the study of atom-resolved charge transport in the Fermi-Hubbard model, the site-resolved observation of molecules, and the creation of bilayer Fermi-Hubbard systems.
\end{abstract}

\maketitle

Understanding strongly correlated quantum systems poses a major challenge both for theory and experiment. Recent years have seen a significant progress in simulating quantum many-body physics with ultracold atoms~\cite{Inguscio2008,bloch2008many,Zwerger2012,Gross2017}. In particular, the Fermi-Hubbard model plays a paradigmatic role in the study of strongly correlated fermions, most prominently for understanding high-$T_c$ superconductivity~\cite{Lee2006doping}. Quantum gas microscopes~\cite{Gross2017,Bakr2009,Sherson2010} of fermionic atoms~\cite{Cheuk2015quantum,Haller2015single,Parsons2015,Omran2015,Edge2015} provide the ability to explore fermion correlations with single-atom, single-site resolution. Recent works have demonstrated the metal and Mott insulator crossover~\cite{Cocchi2016equation,Greif2016,cheuk16Mott,Drewes2016thermodynamics}, studied spin and charge correlations~\cite{Parsons2016, Boll2016, Cheuk2016a, Drewes2016Antiferro,Mazurenko2017,Brown2017}, revealed magnetic polarons~\cite{Koepsell2019} and studied spin~\cite{Nichols2019}, charge~\cite{Brown2019} and heat transport~\cite{GuardadoSanchez2020}. However, most experiments employ fluorescence imaging directly on the lattice used for Hubbard physics. Light-assisted collisions then remove atom pairs residing on the same lattice site from the image~\cite{DePue1999, Schlosser2001}, leading to parity projection~\cite{Bakr2009, Sherson2010} and in particular the appearance of doubly occupied sites~(doublons) as holes. Such Fermi gas microscopes thus measure only the density of singly occupied sites~(singlons), i.e. the local moment~\cite{Cheuk2016a}. The full density can be obtained via absorption imaging~\cite{Cocchi2016equation} but without single site resolution, or by selectively imaging either singlons or doublons~\cite{Mitra2018}.

\begin{figure}[!t]
    \centering
	\includegraphics[width=\columnwidth]{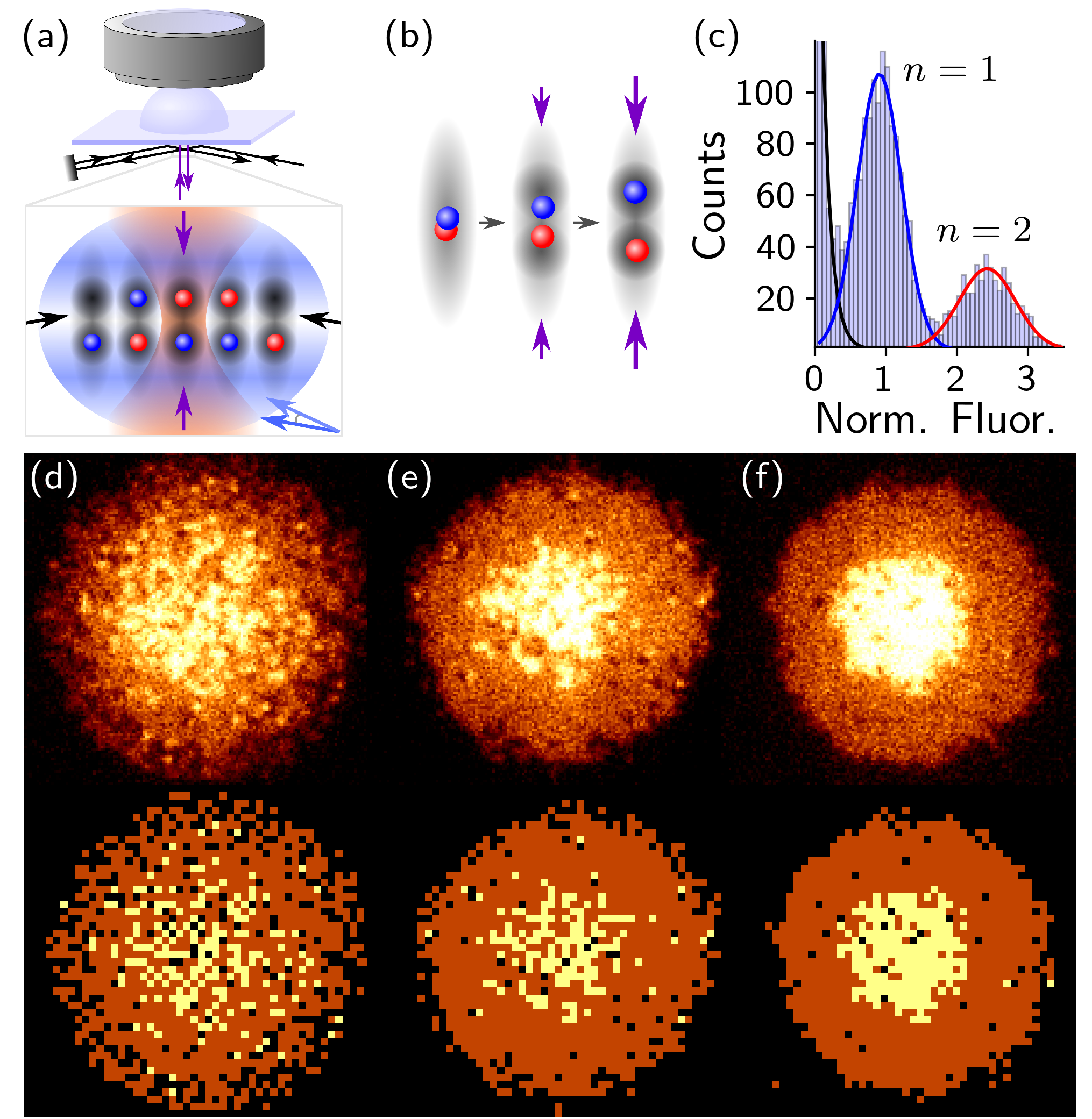}
	\caption{
	{\bf Observing the formation of fermionic Mott and band insulators via total density readout in a bilayer microscope.} (a) A degenerate Fermi gas is prepared in a 2D optical lattice potential (black arrows) beneath a microscope objective. A vertical superlattice~(purple arrows, $532$~nm separation) can hold two atoms in different layers simultaneously within the microscope focus~(collecting $770$~nm light, orange shading). The intensity of Raman light~(blue arrow and shading) used for imaging is tunable for each layer by changing the beam angle. (b) Repulsively interacting atom pairs, originally in a single well, are split by imposing the vertical superlattice before imaging. (c) A typical fluorescence histogram, clearly indicating the presence of $n{=}1$ and $n{=}2$ atoms per lattice site. 
	(d-f) A strongly correlated metal ($U/t{\sim}7$) turns into a fermionic Mott insulator (with $n{=}1$) surrounding a band insulator ($n{=}2$) upon increasing $U/t$ to (e)~$19$ and (f)~$84$. Reconstructed lattice occupations shown below.
	}
	\label{fig:1}
\end{figure}

Revealing the microscopic correlations giving rise to macroscopic observables of the Fermi-Hubbard model requires single-shot measurements of the full density. As the prime example, the fluctuation-dissipation theorem~\cite{zhou2011universal} relates the compressibility to the global number fluctuations of the system via the temperature, requiring measurements of the total density sensitive to atomic shot noise~\cite{Gemelke2009,Sanner2010, Muller2010}. The importance of non-local density fluctuations has been demonstrated~\cite{Drewes2016thermodynamics, cocchi2017measuring}, but revealing their microscopic origin requires site-resolved density measurements.

Progress in fluorescence imaging of the total density was achieved with superlattices~\cite{Boll2016} that spatially separated atom pairs into distinct wells, revealing the interplay of charge and spin~\cite{Koepsell2019} in systems of $\sim 6{\times} 6$~sites.

In this Letter, we introduce a bilayer Fermi gas microscope enabling full site-resolved density readout of large (${\sim} 1500$ sites) 2D Fermi-Hubbard systems in a single fluorescence image. This directly yields the equation of state as pressure, compressibility and doublon density are obtained as a function of density. Site-resolved density correlations reveal the importance of non-local correlations, from the Pauli hole at low filling to strong doublon-hole correlations at half filling. The measured density fluctuation and compressibility directly yield a theory-independent thermometer via the fluctuation-dissipation theorem~\cite{zhou2011universal}. In the Mott insulator, we find strongly correlated nearest-neighbor doublon-hole pairs, required to compensate local density fluctuations to yield the near-vanishing compressibility.


To record the full density information, our setup consists of a bilayer optical lattice potential beneath a microscope objective, shown schematically in Fig.~\ref{fig:1}(a). In the experiment, a 2D Fermi-Hubbard gas is prepared in a single horizontal layer of a 3D optical lattice as reported in~\cite{Cheuk2015quantum}, with horizontal (vertical) lattice spacing of $a{=}541\,\rm nm$ (3~$\mu$m).
For imaging, the depth of the horizontal lattices is increased to prevent tunneling in the 2D plane. Some lattice sites will contain doublons.
We now impose a vertical superlattice~(purple arrow in Fig.~\ref{fig:1}(a)) with $532$~nm spacing, created by retro-reflecting a 1064~nm laser beam off the flat surface of the hemispheric microscope objective.
Driven by Feshbach enhanced repulsive interactions, two atoms originally in a single lattice site separate vertically into different wells~(Fig.~\ref{fig:1}(b))~\cite{SI}.
After the splitting process, Raman sideband cooling is performed as in~\cite{Cheuk2015quantum} and emitted optical pumping photons are collected through the microscope objective.
In contrast to previous work with bosons~\cite{Preiss2015}, the layer separation is within the depth of focus of the microscope, allowing atoms in both layers to be simultaneously imaged onto the same diffraction limited spot on the camera. 

We now demonstrate that separated atoms continue to fluoresce without light-induced loss. By raising the harmonic trapping potential, we create a band insulator at the center of the cloud and perform the vertical separation of atom pairs before imaging. Fig.~\ref{fig:1}(c) shows a typical histogram of an image, with fluorescence counts from singly occupied sites clearly distinguishable from those for originally doubly occupied sites. The fluorescence obtained from atoms in each layer can be tuned via the intensity of Raman light~\cite{SI}. 
Typical images are shown in Fig.~\ref{fig:1}(d-f), for various values of the ratio $U/t$ between the on-site interaction strength $U$, and the tunneling rate $t$~\cite{cheuk16Mott}.
Fig.~\ref{fig:1}(d) shows a strongly correlated metal at $U/t{\sim}7$. As $U/t$ is increased, the tell-tale ``wedding cake'' structure emerges, with a central band insulator at a fluorescence level corresponding to $n{=}2$ surrounded by a Mott insulator at lower fluorescence corresponding to $n{=}1$. Singly and doubly occupied sites are clearly distinguished~\cite{SI}, leading to the digitized images below.

\begin{figure}[t]
	\centering
	\includegraphics[width=\columnwidth]{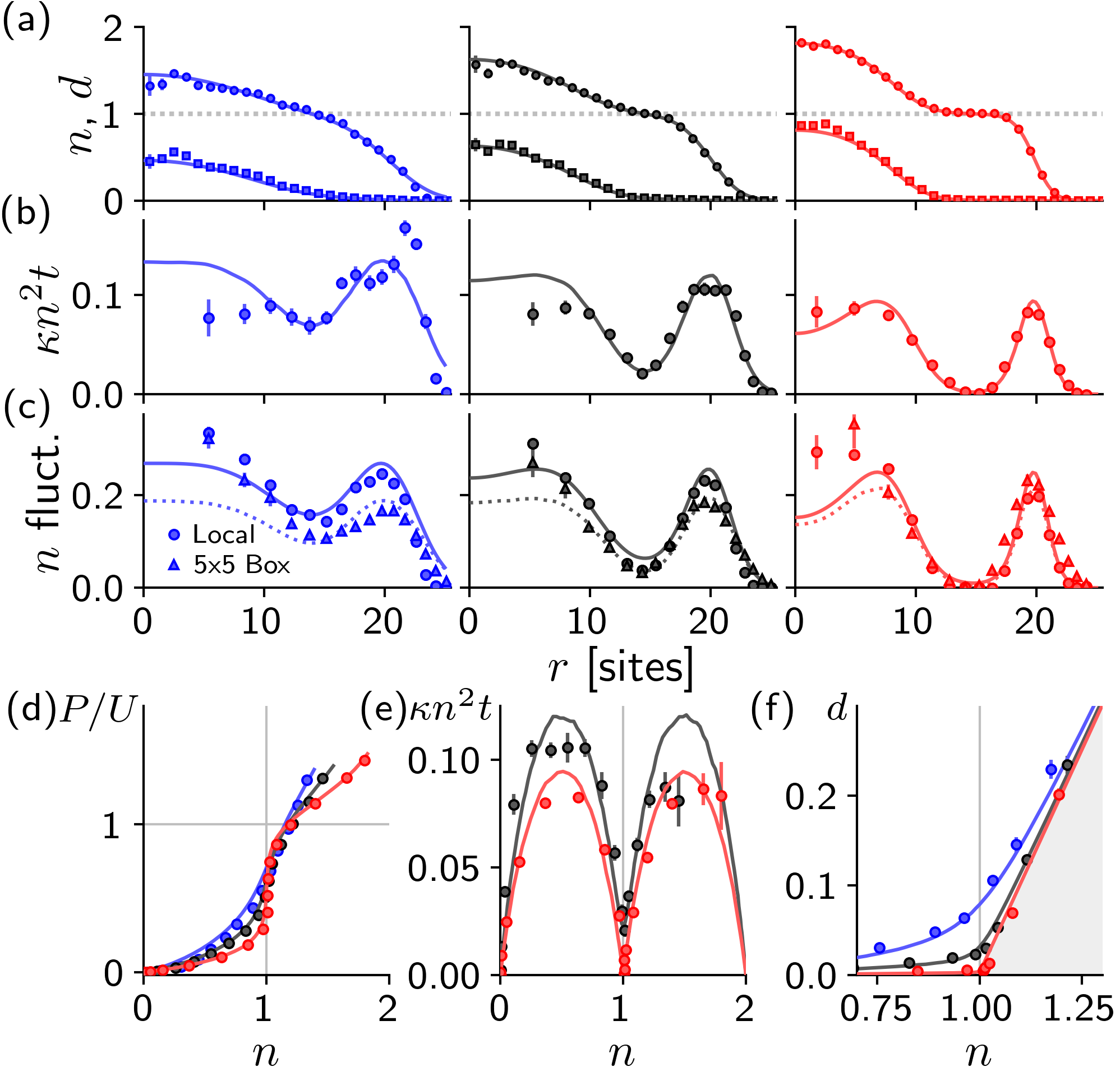}
	\caption{{\bf Equation of State of the 2D Fermi-Hubbard model.} 
	(a) Radially averaged profiles of total density (circles) and doublon density (squares) in a Fermi-Hubbard gas at $U/t=7.1(4)$~(blue), $11.8(5)$~(black), and $25.3(6)$~(red). (b) Measured normalized compressibility $\kappa n^2 t$. (c) Local density fluctuations $\langle \hat{n}^2\rangle - \langle \hat{n}\rangle^2$ (circles) and total atom number fluctuations per area $(\langle \hat{N}^2\rangle - \langle \hat{N}\rangle^2)/$Area in a $5{\times} 5$ box (triangles).
	(d-f) Thermodynamic variables vs.~density: (d) normalized pressure $P/U$, (e) compressibility $\kappa n^2 t$, and (f) doublon density $d$. All lines show Monte Carlo predictions~\cite{varney2009quantum} for $T/t=1.4$ (blue), $T/t=1.6$ (black), and $T/t=2.25$ (red) with the same $U/t$ as the data. Here and elsewhere, data are corrected by measured rates of loss ($\sim 5\%$) and hopping ($\sim 5\%$)~\cite{SI}. 
	}
	\label{fig:2}
\end{figure}

Fig.~\ref{fig:2}(a) shows examples of radially averaged density $n$~(circles) and doublon density $d$~(squares) at varying $U/t$. On a given lattice site, we set $d{=}1$ when $n{=}2$, and the hole density $h{=}1$ when $n{=}0$. With increasing repulsion (from left to right) a Mott plateau emerges at $n {=} 1$. The compressibility $\kappa$ in Fig.~\ref{fig:2}(b) is obtained via the local density approximation from the variation in the measured local potential $V(r)$ as $\kappa n^2 = \left. \partial n/\partial \mu \right|_T = -\left. \partial n/\partial V \right|_T$~\cite{Ku2012,zwierlein2016thermodynamics}. It is observed to vanish in the region of the Mott plateau, directly indicating insulating behavior~\cite{Duarte2015compressibility,Cocchi2016equation}. A simultaneous reduction in local (on-site) fluctuations in the density in Fig.~\ref{fig:2}(c), $\langle \hat{n}^2 \rangle - \langle \hat{n} \rangle^2 = n(1-n)+2 d$, is caused by the reduced double occupancy $d$ in the Mott insulator at $n{=}1$.

Access to the total density directly yields a measurement of the equation of state of the Fermi-Hubbard model.
The canonical equation of state relates pressure $P = P(n, T, U, t)$ to density, temperature $T$ and interaction parameters $U$ and $t$. However, one is free to replace e.g.~temperature by any other thermodynamic variable like the doublon fraction, and e.g.~$t$ by compressibility $\kappa$, thereby obtaining an equation of state of directly and locally observable quantities~\cite{Ku2012,zwierlein2016thermodynamics}.
From the variation of density with potential $n(V)$ one obtains the pressure $P(V)= \int_{-\infty}^{\mu_0-V} n(\mu')\,{\rm d}\mu' = \int_{V}^{\infty} n(V')\,{\rm d}V'$~\cite{Cheng2007trapped, Ho2010obtaining,Nascimbene2010,Ku2012,cocchi2017measuring}. Together with $P$, one has the compressibility $\kappa n^2 = n \left. \partial n/\partial P\right|_T$, and the dimensionless doublon fraction $d$, all as a function of density $n$ (Fig.~\ref{fig:2}(d-f), respectively).
For the strongest interactions it can be observed how the pressure needs to rise above $U$ before breakdown of the Mott insulator occurs and the density can grow above $n{=}1$. 
Finally, the compressibility, together with the total density fluctuations in Fig.~\ref{fig:2}(c) directly yield the temperature $T$ via the fluctuation-dissipation theorem. To this end, in the following we will investigate density correlations.

\begin{figure}[t]
	\centering
	\includegraphics[width=\columnwidth]{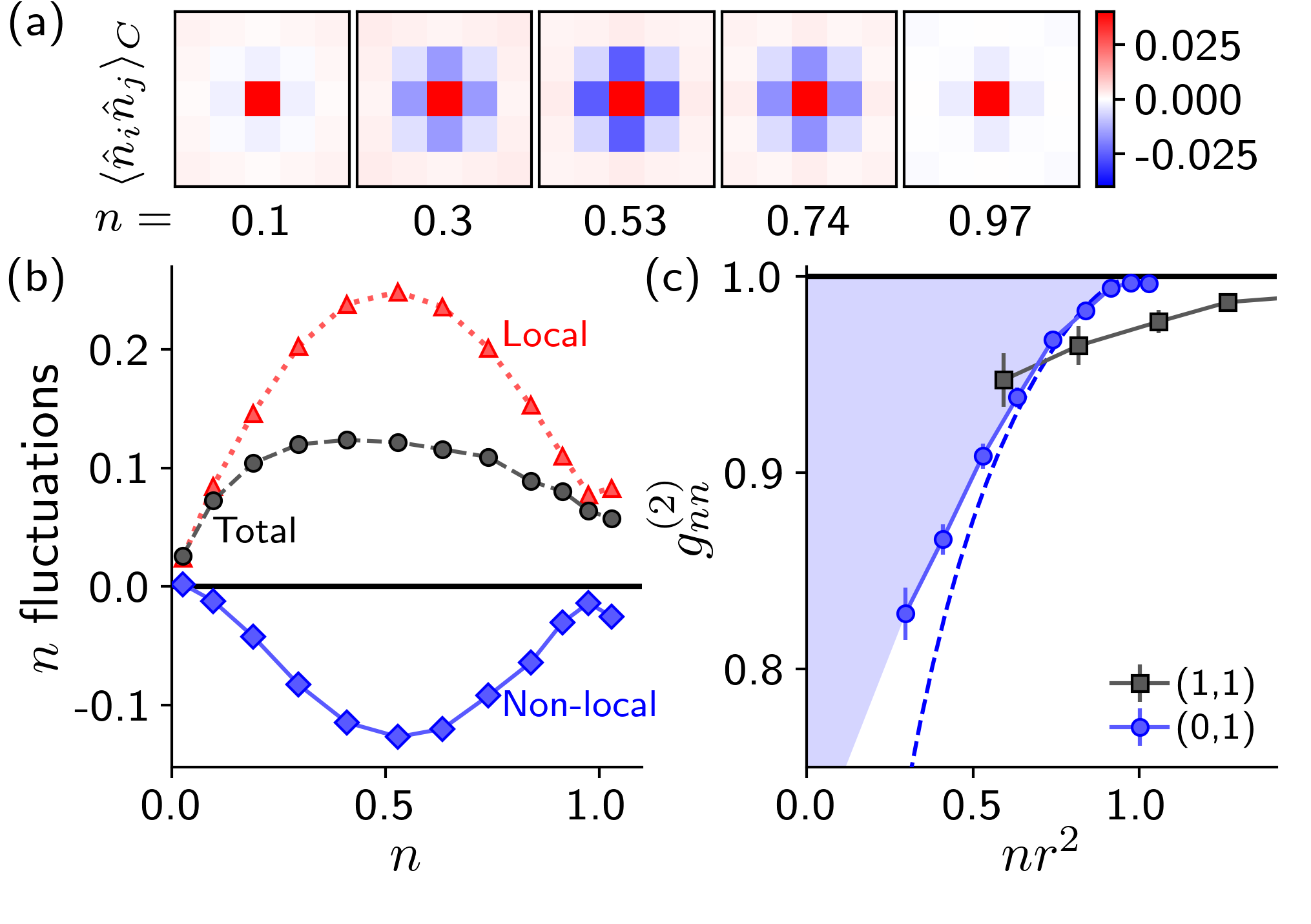}
	\caption{{\bf Measurement of non-local density correlations in the 2D Fermi-Hubbard model.}
	(a) Connected density-density correlations at various densities at $U/t = 11.8(5)$. 
	(b) Density fluctuations $\sum_\delta\langle \hat{n}_i\hat{n}_{i+\delta}\rangle_C$ (total, black circles), $\langle \hat{n}^2\rangle - \langle \hat{n}\rangle^2$ (local, red triangles), and $\sum_{\delta\neq 0}\langle \hat{n}_i\hat{n}_{i+\delta}\rangle_C$ (non-local, blue diamonds). (c) Density-density correlation function $g^{(2)}_{nn}$ for displacements $(0,1)$ (blue circles) and $(1,1)$ (black squares) vs.~$n r^2$, and theory for a non-interacting single-component Fermi lattice gas for displacement $(0,1)$ (blue dashed line) at $T/t = 0.69$ (consistent with thermometry in Fig.~\ref{fig:4}). The shading is a guide to the eye indicating the correlation hole due to Pauli exclusion between like spins and repulsion between unlike spins.
	}
	\label{fig:3}
\end{figure}

The density correlations of a non-interacting Fermi gas are determined by Pauli exclusion, which forbids two identical fermions to share the same phase-space cell. At non-degenerate temperatures, the probability to find two like fermions near each other is suppressed for distances smaller than the thermal de Broglie wavelength $\lambda_{\rm dB}\sim a\sqrt{t/T}$. As the phase space density $n \lambda_{\rm dB}^2/a^2 \gtrsim 1$, i.e. $T \lesssim n\,t$, the size of this Pauli exclusion hole saturates to the spacing $a/\sqrt{n}$ between identical fermions.
In a two-state mixture of fermions and at low filling, repulsion between unlike spins further deepens the correlation hole between particles. These non-local anti-correlations have the effect of reducing the total atom number fluctuations in a given region. Any local upward density fluctuation will be partially compensated by a reduction in nearby density. In Fig.~\ref{fig:2}(c) we demonstrate that density fluctuations are reduced in a $5{\times}5$ site box (triangles) compared to onsite fluctuations (circles), indicating the presence of non-local anti-correlations between fermions. 

We now use the full site-resolved density read-out of our microscope to directly measure the correlation hole in an interacting Fermi-Hubbard lattice gas. The Pauli hole has been inferred from antibunching of the parity-projected density in previous work~\cite{Cheuk2016a}. The connected density-density correlation $\langle\hat{n}_i\hat{n}_{i+\delta} \rangle_C =\langle\hat{n}_i\hat{n}_{i+\delta}\rangle -n_i n_{i+\delta} $ characterizes the non-trivial correlation of finding two particles a distance of $\delta$ lattice sites apart, beyond that for uncorrelated particles at the same density. Fig.~\ref{fig:3}(a) shows the spatial dependence of $\expval{\hat{n}_i\hat{n}_{i+\delta}}_C$ at various densities. Strong non-local anti-correlations are clearly visible.
Fig.~\ref{fig:3}(b) reports the total, local, and non-local density fluctuations. Significant negative non-local correlations indicate a de Broglie wavelength which extends over multiple lattice sites, requiring $T \sim t$~\cite{Walsh2019critical,Kim2019spin}. We note that non-local correlations were inferred but not directly measured in~\cite{Drewes2016thermodynamics}.
The magnitude of local and non-local fluctuations is maximal at $n{\approx}0.5$, a direct consequence of strong on-site repulsion between unlike spins. This effectively reduces the available area for each species by half. On-site density fluctuations are thus equal to that of a single spin species in half the area, of density $n$ and binomial fluctuation $\langle \hat{n}^2\rangle - \langle \hat{n}\rangle^2 \approx n(1-n)$, peaking at $n{=}0.5$. Pauli exclusion requires a corresponding anti-correlation in the area surrounding a given local fluctuation, so non-local fluctuations peak near the same filling. 

The spatial Pauli hole is directly visualized through the density-density correlation function $g^{(2)}_{nn} = \langle\hat{n}_i\hat{n}_{i+\delta}\rangle /n_i n_{i+\delta}$. Fig.~\ref{fig:3}(c) shows the measured $g^{(2)}_{nn}$ for nearest-neighbor and next-nearest neighbor displacements $\delta$ versus $n r^2$, which normalizes distance by the Fermi wavelength. The strong reduction of $g^{(2)}_{nn}$ within one interparticle spacing~(blue shaded region) represents the direct observation of the correlation hole due to Pauli exclusion of like spins, and repulsion of unlike spins. The $g^{(2)}_{nn}$ for a single, non-interacting fermionic species at the full density $n$ shows good agreement, highlighting again that strong interspin repulsion reduces the available area for a given spin species by half. 

\begin{figure}[t]
	\centering
	\includegraphics[width=\columnwidth]{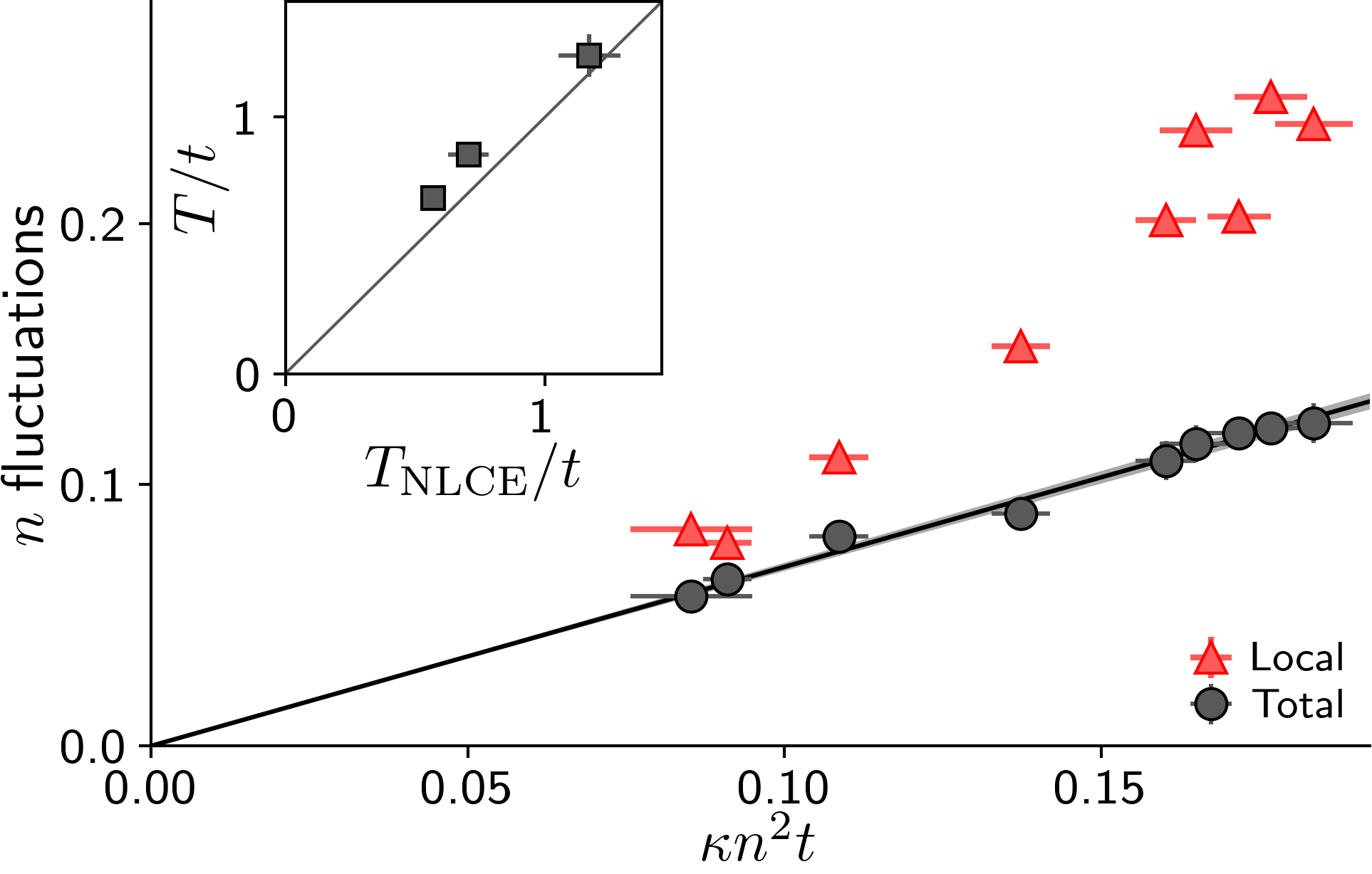}
	\caption{
	{\bf Direct thermometry via density-density correlations.} 
	Density fluctuations vs.~normalized compressibility $\kappa n^2 t$ for $U/t=11.8(5)$:  local fluctuations $\langle \hat{n}^2\rangle - \langle \hat{n}\rangle^2$
	(red triangles) and total fluctuations $\sum_{\delta} \expval{\hat{n}_i\hat{n}_{i + \delta}}_C$ (black circles). 
	A linear fit of total fluctuations vs.~compressibility, fixed through the origin (black solid line), provides the temperature $T = \sum_{\delta} \expval{\hat{n}_i\hat{n}_{i + \delta}}_C/n^2 \kappa$ (grey shading shows statistical uncertainty). Inset:~Measured temperatures vs.~inferred temperatures $T_{\rm NLCE}$ from theoretical fits to radial singlon profiles, after heating the system for variable time~\cite{SI}. Errors are only statistical.
	}
	\label{fig:4}
\end{figure}

With access to both the measured microscopic density fluctuations~(Fig.~\ref{fig:3}) and the macroscopic compressibility~(Fig.~\ref{fig:2}), we are now in the position to probe the fundamental correspondence between fluctuations and response in thermal quantum systems~\cite{Callen1951}. The general density fluctuation-dissipation theorem 
\begin{equation} 
    \kappa n^2 = \left.\frac{\partial n_i}{\partial \mu} \right|_T = \beta \sum_{\delta} \expval{\hat{n}_i\hat{n}_{i + \delta}}_C,
    \label{eqn:FlucDis}
\end{equation}
where $\beta=1/T$, relates directly measurable macroscopic and microscopic quantities without reference to any theoretical model~\cite{zhou2011universal}. Significantly, non-local density correlations will remain a sensitive thermometer down to $T{=}0$ for any compressible system because $\kappa n^2=\left.\partial n_i/\partial \mu \right|_T$ will saturate to the density of states at low temperatures~\cite{Sanner2010}. E.g.~in low density metallic regions with a free particle energy dispersion, $\left.\partial \mu/\partial n_i \right|_T\to \, 2 \pi t$ as $T{\to}\, 0$, which implies sensitivity to temperatures $T\ll t$.
Moreover, by averaging over the system's area, Eqn.~\eqref{eqn:FlucDis} relates compressibility to the global atom number fluctuations: $\kappa n^2= \beta (\langle \hat{N}^2\rangle - \langle \hat{N}\rangle^2)$/Area. 
In small subsystems, however, number fluctuations are enhanced due to non-local correlations across boundaries.
This is the origin of the violation of the area law for entanglement entropy already present for non-interacting fermions~\cite{Wolf2006Fermiarealaw,Gioev2006Fermiarealaw,Swingle2010Fermiarealaw}.

Fig.~\ref{fig:4} shows the total connected density-density correlation~(black circles) versus the normalized compressiblity~$\kappa n^2 t$ for the same dataset as Fig.~\ref{fig:3}.
A linear fit results in a temperature of the cloud of $T/t = 0.69(2)$ using Eqn.~\eqref{eqn:FlucDis}. The entire inhomogeneous atomic gas contributes data, providing high statistical precision for this single parameter fit. Moreover,
the agreement of the data with a linear fit demonstrates that any individual measurement realizes a spatially localized thermometer. For comparison, local fluctuations~(red triangles) are non-linear and are consistently larger than total fluctuations, highlighting again the importance of negative non-local correlations, inferred in~\cite{Drewes2016thermodynamics}. 

We quantitatively benchmark the fluctuation thermometer by independently obtaining the temperature from fits of the radial singlon profiles of the same data to numerical linked-cluster expansion (NLCE) calculations~\cite{Khatami2011Thermodynamics}. Thermometry is repeated for different amounts of heating of the atom cloud~\cite{SI, Cheuk2016a}.
As demonstrated in the inset of Fig.~\ref{fig:4}, the temperatures measured via fluctuation thermometry and those obtained from fits to NLCE agree. Note that in general, comparison of measured quantities to theory requires fitting to non-linear and in some cases non-monotonic functions, leading to difficulties in assessing systematic errors.
In contrast, the sole sources of systematic uncertainty in fluctuation thermometry are the calibration of the trap potential, entering linearly into uncertainty in $T$, and measurable errors in the density. We have thus established a theory-independent, precise and sensitive thermometer for interacting lattice fermions.
The method is also ideally suited for homogeneous systems in box potentials~\cite{Nichols2019}, where density fluctuations in the presence of a well-calibrated linear gradient will provide access to the local temperature. This opens up prospects for the study of heat transport in the Fermi-Hubbard model.

\begin{figure}[t]
	\centering
	\includegraphics[width=\columnwidth]{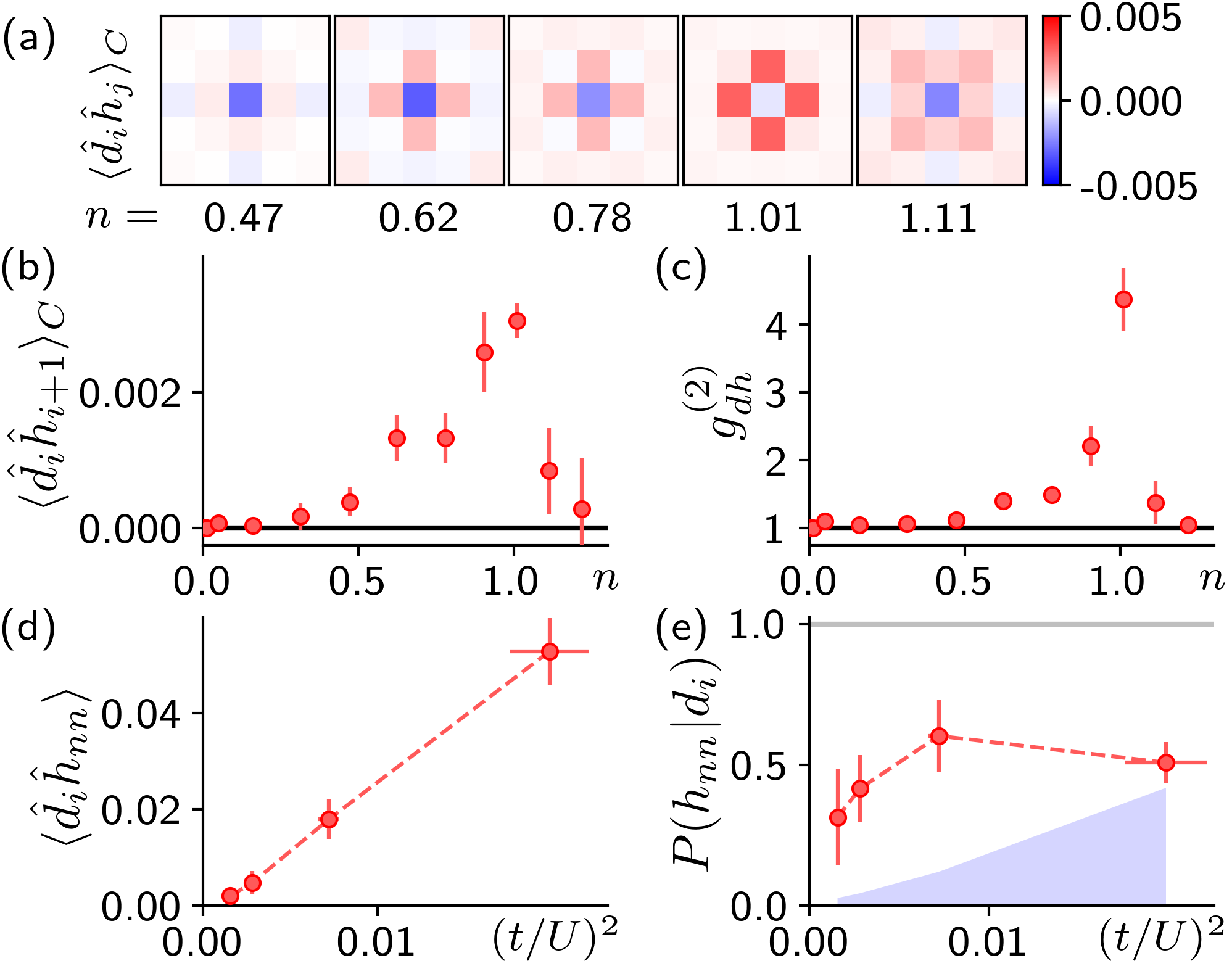}
	\caption{
	{\bf Direct observation of doublon-hole correlations in the 2D Fermi-Hubbard model.} (a) Spatial dependence of $\langle \hat{d}_i\hat{h}_{j}\rangle_C$ for  $U/t = 11.8(5)$ vs.~filling $n$. (b) Nearest neighbor correlations $\langle \hat{d}_i\hat{h}_{i+1}\rangle_C$ and (c) nearest neighbor $g^{(2)}_{dh}$ vs.~$n$. (d) Density of doublons with any nearest neighbor hole pair  $\langle\hat{d}_i\hat{h}_{nn} \rangle = \sum_{j\in nn}\langle\hat{d}_i\hat{h}_{j} \rangle$ at half filling vs.~$(t/U)^2$ (for $U/t = 25.3(6)$, $18.8(5)$, $11.8(5)$, and $7.1(4)$).
	(e) Conditional probability of a nearest neighbor hole $P(h_{nn}|d_i) = \langle\hat{d}_i\hat{h}_{nn} \rangle/d_i$ at half filling compared to the expectation for a random distribution (blue shaded region) at the same hole and doublon density: $4d$.
	}
	\label{fig:5}
\end{figure}

The fluctuation-dissipation theorem provides insight into charge fluctuations in the Mott insulator at half-filling, at temperatures $T \ll U$, where the compressibility vanishes. In any system where either $T \to 0$ or $\kappa n^2 \to 0$, Eqn.~\eqref{eqn:FlucDis} implies that local and non-local density fluctuations must cancel. For finite tunneling $t \sim T \ll U$, the system remains insulating, although the local operator $t$ acts as a perturbation that causes charge fluctuations over short distances~\cite{cocchi2017measuring}. The dominant contributions to $\langle \hat{n}_i\hat{n}_{j}\rangle_C = \langle \hat{d}_i\hat{d}_{j}\rangle_C + \langle \hat{h}_i\hat{h}_{j}\rangle_C -2\langle \hat{d}_i\hat{h}_{j}\rangle_C$ are nearest neighbor doublon-hole fluctuations which occur with probability $\sim (t/U)^2$~\cite{Endres2013single}. Their existence has been inferred in~\cite{Cheuk2016a} by observing bunching of holes after parity projection. For fermions, these nearest neighbor doublon-hole correlations signal spin singlet formation, as Pauli exclusion prevents tunneling for spin triplets. 

Armed with full density read-out, in Fig.~\ref{fig:5} we now directly detect these doublon-hole fluctuations. At our temperatures $T \ll U$, where thermal fluctuations are frozen out, doublon-hole fluctuations are purely quantum in origin. 
Many of these isolated doublon-hole pairs can be directly observed as fluctuations within the strongly coupled Mott insulator in
Fig.~\ref{fig:1}(e)  $(U/t = 18.8(5))$.
In Fig.~\ref{fig:5}(a-c) we show the spatial dependence of the connected doublon-hole correlator $\langle\hat{d}_i\hat{h}_{j} \rangle_C$, the nearest neighbor correlator $\langle\hat{d}_i\hat{h}_{i+1} \rangle_C$, and the doublon-hole distribution function $g^{(2)}_{dh} = \langle\hat{d}_i\hat{h}_{i+1} \rangle/d_i h_{i+1}$ versus density at $U/t = 11.8(5)$, all of which demonstrate strongly enhanced local doublon-hole correlations near $n {=}1$. 

In  Fig.~\ref{fig:5}(d) we report the nearest neighbor doublon-hole pair density  $\langle\hat{d}_i\hat{h}_{nn} \rangle = \sum_{j\in nn}\langle\hat{d}_i\hat{h}_{j} \rangle$ with respect to $(t/U)^2$. The linear relationship highlights the physical origin of doublon-hole pair correlations in a coherent, off-resonant tunneling process of amplitude $\sim t/U$.
To demonstrate the strength of bunching, we obtain the conditional probability $P(h_{nn}|d_i) = \langle\hat{d}_i\hat{h}_{nn} \rangle/d_i$ to find a hole next to a doublon in Fig.~\ref{fig:5}(e). As a comparison, we also show the conditional probability for a Poisson process at the same hole and doublon density $4 d$~(blue shaded area). At small $t/U$, the conditional probability far exceeds random chance, showing that doublons and holes are tightly bound in a Mott insulator.

In conclusion, we demonstrate a robust method to measure the total site-resolved density in a cold-atom realization of the 2D Fermi-Hubbard model. We use this ability to directly detect non-local correlations, in particular the Pauli correlation hole at low filling and doublon-hole correlations in the Mott insulating region. Model-free thermometry is established via the fluctuation-dissipation theorem. Using a magnetic field gradient, we can also perform spin dependent splitting~\cite{SI}, which will eventually allow simultaneous observation of both charge and spin. Our superlattice geometry opens up the ability to study bilayer and even multilayer Fermi-Hubbard models, relevant for high-temperature superconductivity~\cite{Bulut1992bilayer,Okamoto2008superlattice}.

{\it Note added:} After completion of our experimental work~\cite{DAMOP2019}, a spin-resolved bilayer imaging technique was realized in~\cite{koepsell2020robust}.

We would like to thank M.~A.~Nichols and H.~Zhang for early contributions to the experiment, E. Kozik for stimulating discussions, and E. Khatami for providing NLCE calculations~\cite{Cheuk2016a}.
This work was supported by the NSF, ONR, an AFOSR MURI on “Exotic Phases of Matter”, the David and Lucile Packard Foundation, the Gordon and Betty Moore Foundation through grant GBMF5279, and the Vannevar Bush Faculty Fellowship. M.Z. acknowledges support from the Alexander von Humboldt Foundation.


\setcounter{equation}{0}
\setcounter{figure}{0}
\setcounter{secnumdepth}{2}
\renewcommand{\theequation}{S\arabic{equation}}
\renewcommand{\thefigure}{S\arabic{figure}}
\renewcommand{\tocname}{Supplementary Materials}
\renewcommand{\appendixname}{Supplement}


\section*{Supplementary Materials}

\subsection{Preparation and imaging of doubly-occupied sites}

A typical experimental protocol prepares a gas of $^{40}$K fermions in a balanced mixture of two different hyperfine states ($|1\rangle \equiv |F{=}9/2,m_F{=}-9/2\rangle$ and $|2\rangle \equiv|9/2,-7/2\rangle$) in a single 2D layer of a 3D optical lattice at approximately background repulsive interactions (scattering length $\sim 190~a_0$) near a magnetic field of $151$~G.
Typical lattice potential depths are approximately between $8~E_r$ and $13~E_r$ to balance the effects of tunneling and interactions, where $E_r = (\hbar\pi/\text{541~nm})^2/2m \sim h\times \text{4250~Hz}$ is the horizontal lattice recoil energy. 
To freeze the system for observation, we then ramp the optical lattice depth to $\sim 70~E_r$ in $2.5$~ms which prevents any further in-plane tunneling. Next we prepare the system to separate fermion pairs within a single harmonic well into two different wells of the vertical superlattice. First, we ramp the magnetic field from 151~G to 200~G in 40~ms to enhance the repulsive interactions between fermions by a factor of $\sim 4$ compared to background interactions. In 100~ms we then adiabatically apply a vertical potential gradient to displace the potential minimum of each lattice site vertically to better align with a node of the vertical superlattice, which will subsequently be applied. This vertical potential gradient is formed by an accordion optical lattice of 1064~nm light reflecting off of the microscope substrate at an adjustable angle. If this vertical potential gradient were not applied, the superlattice would be displaced from the single well minimum, leading to an energy offset between the superlattice wells in the deep superlattice limit. The superlattice intensity is then increased in 100~ms from $0~E_r$ to $\sim 300~E_r$ so that the superlattice becomes the dominant vertical potential. During this ramp, the ground state of two fermions in a single well adiabatically maps to the ground state of two fermions in the double well with repulsive interactions: the state with separated fermions. The repulsive fermion interaction energy sets the minimum gap between the ground and excited states during this splitting process, and ensures adiabaticity. Experimentally, it is observed that the procedure is adiabatic and therefore robust to perturbations in timescales, interactions, or applied potentials. The remaining procedure for imaging is identical to procedures described in previous works \cite{Cheuk2015quantum, cheuk16Mott}. We typically collect $\sim 3$~s of Raman sideband cooling optical pumping photons for fluorescence imaging. 

We now describe the discrimination fidelities and loss rates of a typical experiment. A typical fluorescence histogram is shown in Fig.~1 of the main text. This histogram has a doublon detection fidelity $\gtrsim 98.5\%$ and singlon detection fidelity $\gtrsim 92.5\%$, as determined by Gaussian fits to the singlon and doublon histogram peaks with appropriate discrimination thresholds. There is a tradeoff between the intensity of the fluorescence imaging light and the imaging time:  with higher fluorescence one can obtain better doublon-singlon and singlon-hole discrimination fidelities, although loss rates increase. Loss rates are measured as described below. Singlons are not lost during each image with typical measured fidelity $\gtrsim 96\%$, while doublons are not lost (i.e.~not converted into either singlons or holes) with fidelity $\gtrsim 90\%$, which is comparable to twice the singlon loss rate as would be expected for uncorrelated loss of each atom. Hopping rates of atoms in each layer are generally $\lesssim 6\%$.

\subsection{Characterizing simultaneous bilayer imaging}
\begin{figure}[t]
	\centering
	\includegraphics[width=\columnwidth]{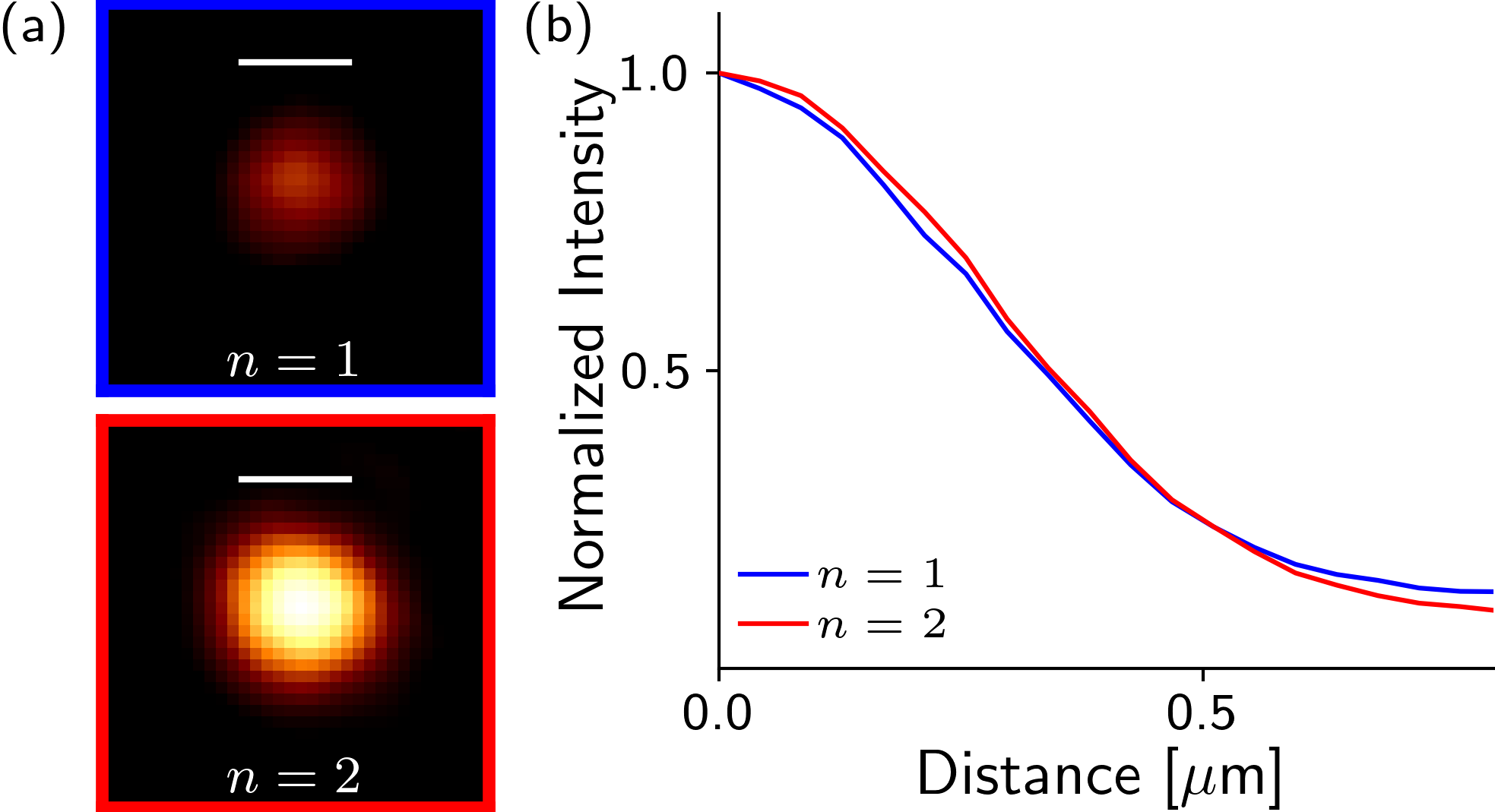}
	\caption{
	(a) Images of isolated originally singly-occupied sites (top) and originally doubly-occupied sites (bottom). Each image is averaged over $\sim 200$~isolated sites. White bar denotes lattice spacing of 541~nm. 
	(b) Radial intensity profiles of (a) normalized to peak intensity, showing that both layers of the microscope are equally in focus. The FWHM of the singlon (doublon) profile is 670~nm (685~nm), which differs by less than 5\% (7\%) from previous work \cite{Cheuk2015quantum}.
	}
	\label{fig:S1}
\end{figure}

In order to measure the full density information of a single-layer Fermi-Hubbard gas, a bilayer quantum gas microscope is used to prevent light-assisted collisions. By using a directly retro-reflected vertical superlattice with wavelength 1064~nm, the distance between the two layers is significantly smaller than similar setups with bosons~\cite{Preiss2015} and fermions~\cite{koepsell2020robust}, allowing us to image both layers simultaneously. The spacing between two superlattice layers~(532~nm) is smaller than the wavelength of the emitted light~(770~nm) during fluorescence imaging, hence atoms in either layer can be in focus at the same time. This removes the necessity of mechanically moving the objective's focus to two different locations, taking two images and referencing them correctly. In Fig.~\ref{fig:S1}(a), we show the intensity profile of isolated lattice sites with one~(blue) or two~(red) atoms averaged over 200 isolated sites. For comparison, the horizontal lattice spacing is depicted as a white bar in each image. The radially averaged intensity profile normalized to the peak intensity for each occupation is shown in Fig.~\ref{fig:S1}(b). The point-spread-functions have comparable shape, which is also similar to previous work~\cite{Cheuk2015quantum}. 

\begin{figure}[t]
	\centering
	\includegraphics[width=\columnwidth]{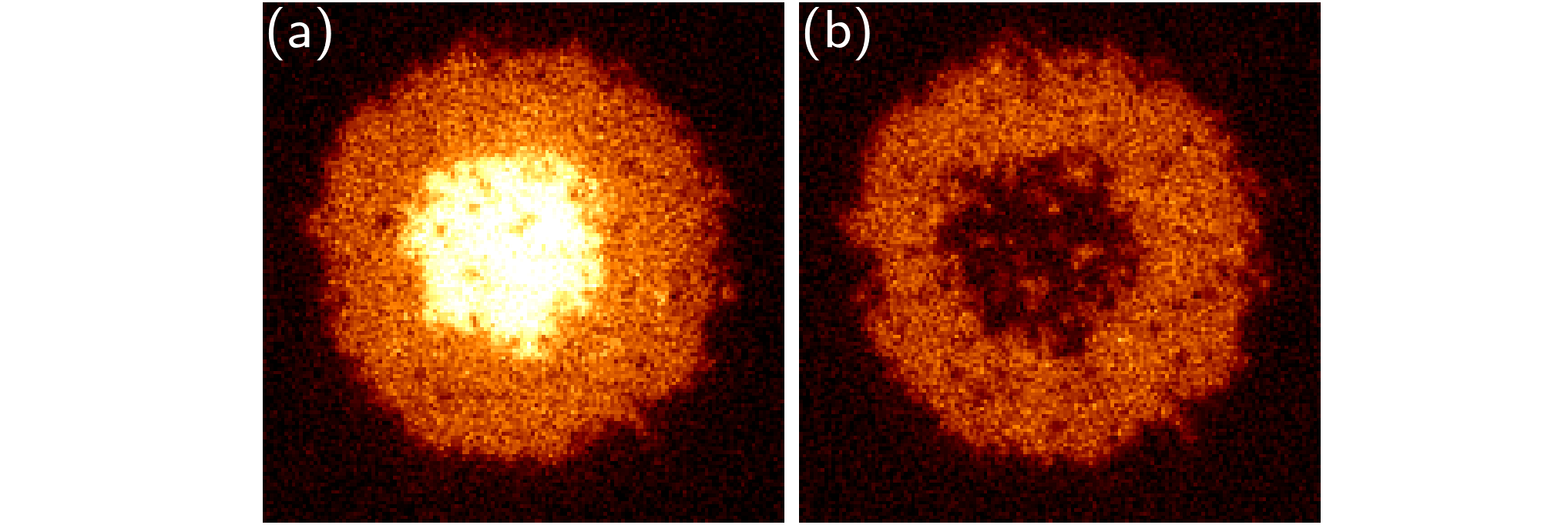}
	\caption{
	(a) Image of the same atomic system as Fig.~1(f) of the manuscript, showing a central region with $n{=}2$ particles per site surrounded by a region with $n{=}1$ particles per site. (b) The same atomic cloud after recombining the two vertical wells of the superlattice so that atom pairs in the central region undergo light-assisted collisions during the second image, thus appearing dark.
	}
	\label{fig:S1p5}
\end{figure}

After acquiring the first image of an atomic cloud (such as Fig.~\ref{fig:S1p5}(a)), and within the same experimental run, we can recombine the atom pairs into a single well again by ramping down the superlattice, and then perform Raman imaging. The resulting image is shown in Fig.~\ref{fig:S1p5}(b), with a dark central region in place of the band insulator in Fig.~\ref{fig:S1p5}(a), reflecting light-assisted collisions ejecting overlapping atom pairs~\cite{DePue1999, Schlosser2001}.

\subsection{Differential fluorescence imaging}

The number of photons obtained from each fluorescing atom is correlated with the local intensity of the Raman beams used in the sideband cooling process during imaging. It is possible to selectively tune the intensity of the Raman light in each layer of the bilayer microscope, which then leads to differential fluorescence of the two layers. 

The horizontally polarized Raman light ($\sim 767$~nm) used in the imaging scheme~\cite{Cheuk2015quantum} is reflected off the substrate at a shallow angle~($\sim 10.8$ degrees), forming an interference lattice with large spacing~($\sim 2$~$\mu$m) in the vertical direction (schematically shown in Fig.~1(a)). By changing the angle at which the beam hits the substrate by less than $0.5$ degrees, we are able to place the interference node at either layer used for imaging ($532$~nm separation) or between the two layers, as measured by the magnitude of fluorescence in each layer. We use a $10$~mm glassplate to repeatably translate the Raman beams by $\sim 1.5$~mm in the Fourier plane, which changes the angle of incidence of the beam on the substrate, and the interference node position. The D1 optical pumping light is independently sent into the chamber at a shallow angle of $\sim 2.5$ degrees. 

\begin{figure}[t]
	\centering
	\includegraphics[width=\columnwidth]{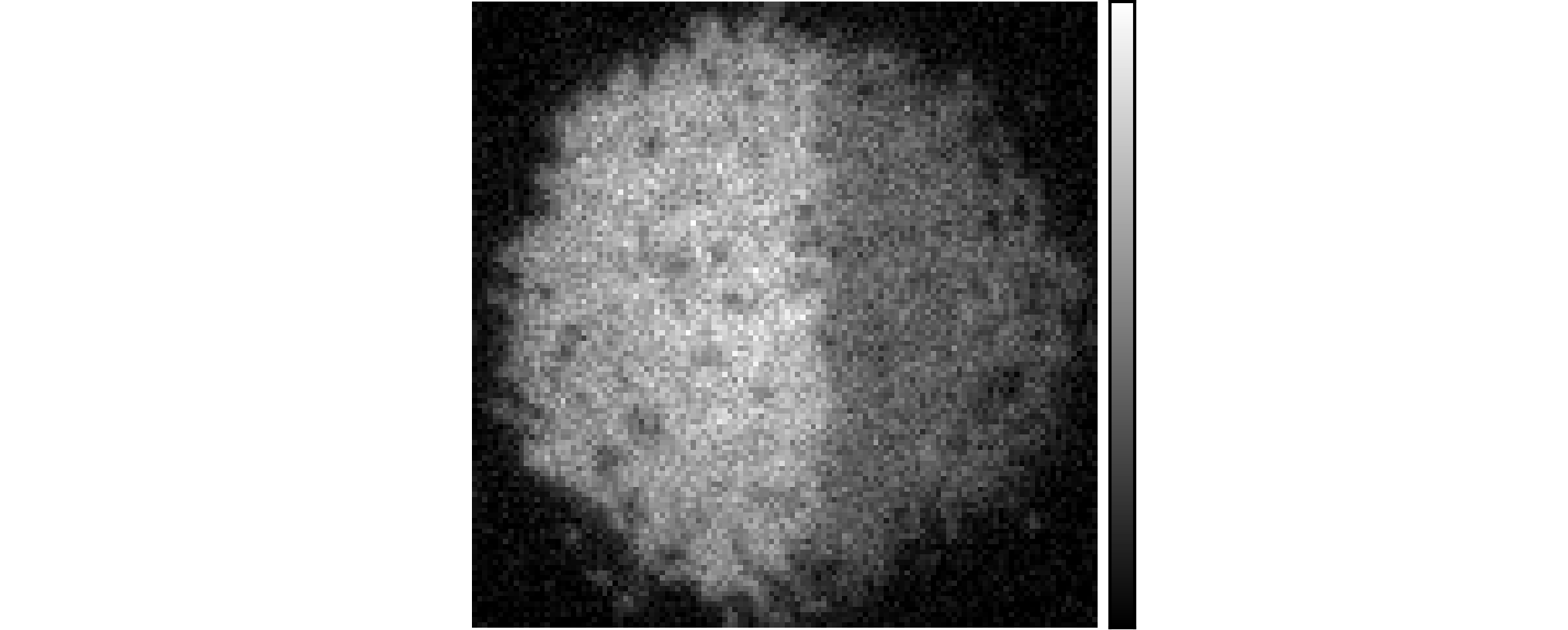}
	\caption{
	Demonstration of differential imaging fluorescence. A vertical optical potential gradient transfers each half of a Mott insulator into different vertical layers of the bilayer microscope before imaging. The intensity of the Raman light is intentionally reduced in one layer.
	}
	\label{fig:S2}
\end{figure}

Fig.~\ref{fig:S2} shows a fluorescence image of a Mott insulator where the left half of the cloud has been placed in one vertical layer for imaging, and the right half has been placed in the other layer. This is achieved via an applied vertical optical potential gradient which varies in strength from left to right. A Raman interference node is closer to the layer on the right half of the image, so that the Raman intensity there is lower.

Reducing the Raman light intensity in a given layer can reduce the fluorescence of that layer significantly without leading to additional heating. The Raman sideband cooling scheme accumulates atoms in a dark state of the optical pumping light and Raman light. However, the Raman light is intentionally spectrally broadened to transfer atoms out of this dark state and lead to fluorescence \cite{Cheuk2015quantum}. With lower Raman intensity each atom spends more time in this dark state, while the effective cooling rate per cycle is not altered. Each layer of the microscope can thus be ``shelved'' at an interference node of the Raman light and preserved for subsequent imaging. Future work will determine whether separate imaging of each layer can be performed sequentially with high fidelity while the other layer is kept dark. 

\subsection{Imaging loss correction}
During imaging, heating due to scattered photons can lead to the ejection of atoms from the system and to enhanced tunneling to nearby sites (detected as hopping of an atom between nearby sites). During each experiment, multiple subsequent images are taken of the same atomic distribution. The average rate of both loss and hopping can then be measured directly from the reconstructed lattice site occupations in subsequent images. 

Loss is modeled as a probability of converting a given site occupation to a lower number of particles (i.e.~a doublon to a singly-occupied site, a doublon to a hole, or a singlon to a hole). Hopping during an image is modeled as a process where two specific states which are present on nearby sites, such as a singlon and a hole, are exchanged with a certain conditional probability $\alpha_{sh}$. Therefore, the rate at which a singlon is converted to a hole due to hopping is proportional to the nearby density of holes $h$, i.e.~is $\alpha_{sh}h$. The overall measured rate of conversion from a singlon to a hole on a specific lattice site then consists of contributions from a constant loss rate, and from hopping processes which depend on the local density, all of which are simultaneously measured. 

Atoms are not lost during imaging with typical measured fidelities of $ \gtrsim 96\%$ for singlon detection, and 
$ \gtrsim 90\%$ for doublon detection. Hopping rates in either layer are typically $\lesssim 6\%$, as inferred from conditional exchange of doublons to singlons, and singlons to holes.  

From the measured loss rates, it is possible to correct for the change in atomic density that occurred while the first image was acquired. This correction involves a simple loss matrix inversion to obtain the real densities $d$, $s$, and $h$ of doublons, singlons, and holes in terms of the observed densities of each type. Hopping processes are not used in the correction, however, as they have no effect on the local density. To infer onsite variances, the formula $n = s + 2d$ is used to obtain $\langle n^2\rangle = s + 4d$ in terms of loss corrected values for $s$ and $d$. When calculating loss corrected correlators such as $\langle \hat{d}_i\hat{h}_{j}\rangle$, a simplifying assumption of independent loss at each site is made so that site $i$ and site $j$ can be  independently corrected for loss. Fractional errors of a few percent will be introduced by this simplifying assumption on measured correlators.


\subsection{Combined spin and charge read-out}
Here we outline how to extend the bilayer microscopy technique to combined spin and charge read-out. The basic principle is to map spin information to spatial information before imaging by transferring all spins of a single species to a specific vertical layer. If it is then possible to extract the atomic distribution in each layer separately, e.g.~by taking two subsequent images of each layer separately using the differential fluorescence imaging technique in Fig.~\ref{fig:S2}, then the full spin and charge information could be obtained. 

Here we report on first steps towards implementing this method. To map each spin to separate layers, a compensating vertical optical potential gradient (as in Fig.~\ref{fig:S2}) is first applied to make it equally likely for either spin to transfer to either well of the vertical double well. Next, because the two spin states $|9/2, -9/2\rangle$ and $|9/2, -7/2\rangle$ have similar magnetic moments, we perform an RF transfer of $|9/2, -9/2\rangle$ to $|7/2, -7/2\rangle$, a state with a large, opposing magnetic moment. We then apply an additional vertical magnetic field gradient to push each hyperfine state in opposite vertical directions before imposing the vertical superlattice. The subsequent adiabatic transfer from a vertical single well to a double well with spin-dependent bias should transfer all singlons of a given spin species to a different, known vertical layer before imaging. 

\begin{figure}[h]
	\centering
	\includegraphics[width=\columnwidth]{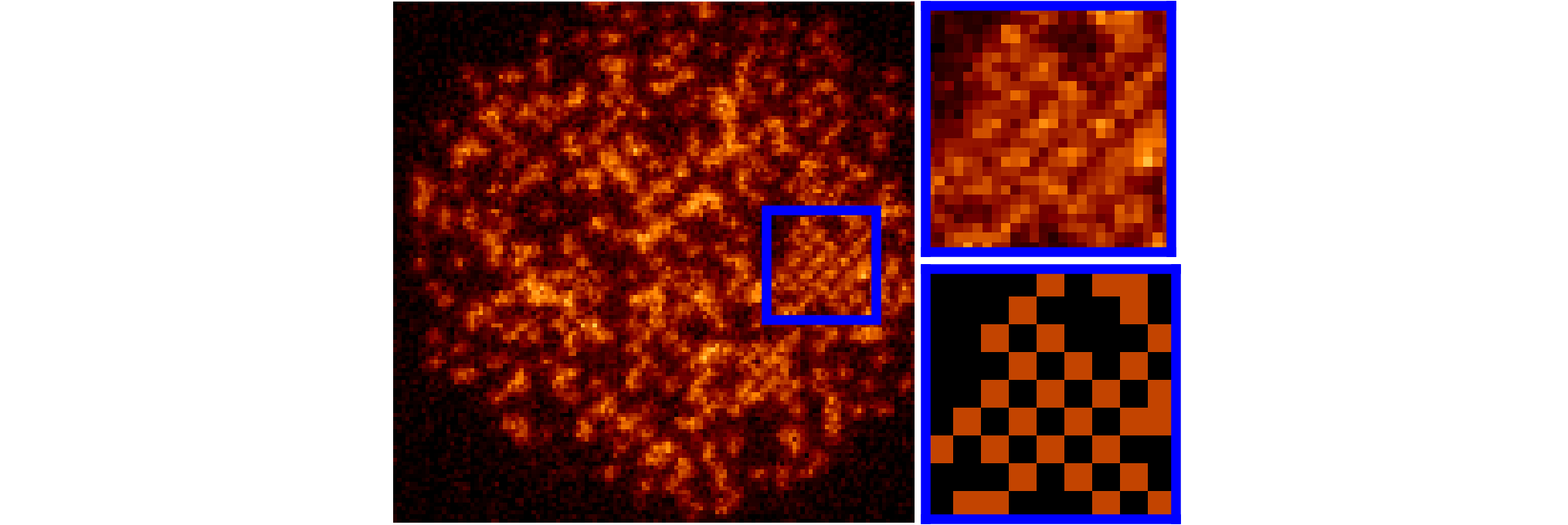}
	\caption{Image of a single spin species that has been transferred to the second vertical layer of the bilayer microscope. 
	}
	\label{fig:S3}
\end{figure}

After performing the above spin-to-spatial information mapping, we image a specific layer using differential fluorescence and we image a specific spin state by removing the other spin species with resonant light, as in \cite{Cheuk2016a}. We confirm that all atoms of either spin are detected only in one layer, and none are detected in the other layer (and reversed for the other spin). In Fig.~\ref{fig:S3}, we show an example image of a single spin species that has been transferred to the second vertical layer via this process. Patches of strong anti-ferromagnetic spin correlations are visible by eye. Combining spin splitting with successive readout of each layer will eventually allow extracting the full spin and charge information. 

\subsection{Fluctuation thermometer sensitivity to heating}
\begin{figure}[h]
	\centering
	\includegraphics[width=\columnwidth]{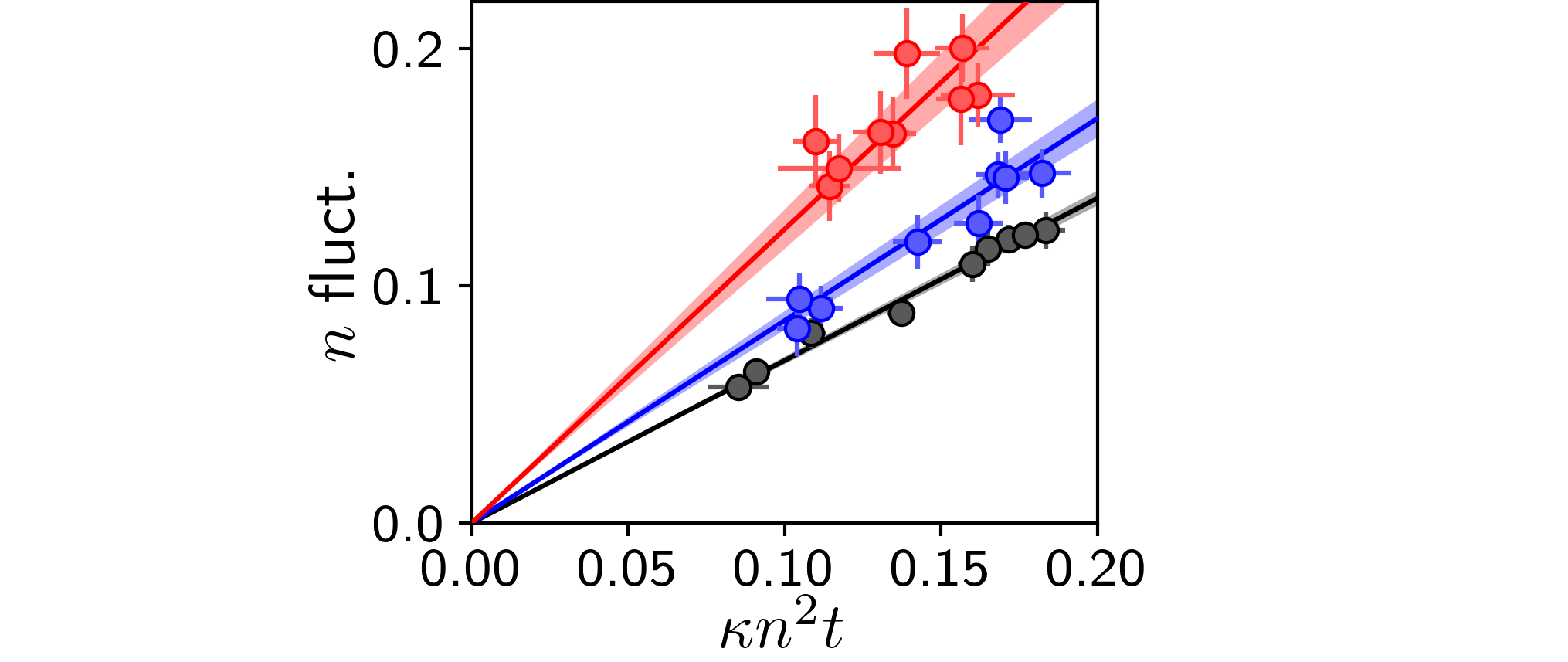}
	\caption{Total connected density-density fluctuations vs.~normalized compressibility for the three measurements in the inset of Fig.~4 of the manuscript, corresponding to heating the system by scattering of lattice photons for 0s, 1s, or 2s (black, blue, or red data) with extracted $T/t=0.69(2)$, $T/t=0.85(4)$, and $T/t=1.24(8)$. Data from ${\sim}70$, ${\sim}35$, and ${\sim}25$ images.
	}
	\label{fig:S4}
\end{figure}

The inset in Fig.~4 of the manuscript demonstrates sensitivity of the fluctuation thermometer to heating. Heating is performed simply by waiting for a given time (up to two seconds), allowing for single-photon scattering from lattice lasers. In Fig.~\ref{fig:S4} we additionally provide the density fluctuations and normalized compressibility which are used to extract the temperatures for each heating time in this inset. By resampling the set of experimental images for each heating time with replacement (bootstrapping), we obtain a best fit temperature (solid lines) and the standard deviation of the fitted temperature (shaded regions).  

\bibliographystyle{apsrev4-1}
\bibliography{BilayerMicroscope}

\end{document}